\documentstyle[12pt]{article}
\input{epsfig.sty}
\newcommand{\beq}{\begin{eqnarray}}
\newcommand{\eeq}{\end{eqnarray}}
\begin{document}
\title{Heavy-quark state production in A-A collisions
 at $\sqrt{s_{pp}}$=200 GeV}
\author{Leonard S. Kisslinger\\
Department of Physics, Carnegie Mellon University, Pittsburgh, PA 15213\\
Ming X. Liu and Patrick McGaughey\\
P-25, Physics Division, Los Alamos National Laboratory, Los Alamos, NM 87545}
\date{}
\maketitle
\begin{abstract}
   We estimate differential rapidity cross sections for
$J/\Psi$ and $\Upsilon(1S)$ production via Cu-Cu and Au-Au collisions at
RHIC, and the relative probabilities of $\Psi'(2S)$ to $J/\Psi$ production
via p-p collisions using our recent theory of mixed heavy quark hybrids, in 
which the $\Psi'(2S)$ mesons have approximately equal normal $q\bar{q}$ and 
hybrid $q\bar{q}g$ components. We also estimate the relative probabilities of 
$\Psi'(2S)$ to $J/\Psi$ production via Cu-Cu and Au-Au collisions, which
will be measured in future RHIC experiments. We also review production ratios
of $\Upsilon(2S)$ and $\Upsilon(3S)$ to $\Upsilon(1S)$ in comparison to recent
experimental results.This is an extension of our recent work on p-p collisions 
for possible tests of the production of Quark-Gluon Plasma via A-A collisions 
at BNL-RHIC.

\end{abstract}
\noindent
PACS Indices:12.38.Aw,13.60.Le,14.40.Lb,14.40Nd
\vspace{1mm}

\vspace{1mm}

\section{Introduction}

  In our recent work on the production of heavy quark states in p-p 
collisions\cite{klm11} we used the color octet model\cite{cl96,bc96,fl96},
which was shown to dominate the color singlet model in studies of $J/\Psi$ 
production at E=200 GeV\cite{nlc03,cln04}. Among other results we found
that our mixed heavy quark hybrid theory\cite{lsk09} correctly predicted the 
experimental result for the ratio of $\Upsilon(3S)$ to $\Upsilon(1S)$ cross 
sections\cite{fermi91}, while the standard model of $|b\bar{b}>$ for the 
$\Upsilon(3S)$ state was about a factor of three too small. Also the recent
measurement of the ratio $\frac{\sigma(\Upsilon(2S))+\sigma(\Upsilon(3S))}
{\sigma(\Upsilon(1S))}$ via p-p collisions by LHC-CMS\cite{cms11} was in 
agreement with the mixed hybrid theory for the $\Upsilon(3S)$, while the 
standard model was more than a factor of two smaller. This will be discussed 
in detail below.

  In our present work we study heavy quark state production at RHIC, with
E=$\sqrt{s_{pp}}$ = 200 GeV for $A-A$ ($A=N+Z$) collisions. Since the present 
BNL-RHIC cannot measure the $\Upsilon(1S), \Upsilon(2S), \Upsilon(3S)$ 
separately, we shall mainly study the $J/\Psi(1S)$ and $\Psi'(2S)$ production. 
In section 2 we briefly
review the mixed hybrid theory for charmonium and bottomonium states,
the color octet model, and the relation between the standard and mixed
hybrid theories of heavy quark meson states needed for our
present work. In section 3 we discuss the production of $J/\Psi$ and 
$\Upsilon(1S)$ states in Cu-Cu and Au-Au collisions, with results based
on the research of many preceding theorists and experimentalists. Note that
due to uncertainty in normalization of the absolute cross sections one main 
prediction is the shapes of the rapidity dependence rather 
than the magnitudes of the cross sections. The other main prediction is the
ratios of cross sections. In
section 4 we discuss the ratio of $\Psi'(2S)$ to $J/\Psi(1S)$ production
and compare the hybrid vs standard theory to recent experiments
with p-p collisions; and predict this ratio for future A-A collision RHIC 
experiments using recent experimental ratios of $Pb-Pb$ to $p-p$ 
$\Upsilon(mS)$ production.
 
\section{Review of mixed hybrid heavy quark mesons and the 
color octet model}

  We give a very brief review of the hybrid heavy quark and color octet
models, and their relationship for the present work. See Ref\cite{klm11}
for details.

  Using the method of QCD sum rules it was shown\cite{lsk09} that
the $\Psi'(2S)$ and $\Upsilon(3S)$ are approximately 50-50 mixtures
of standard quarkonium and hybrid quarkonium states:
\beq
        |\Psi'(2s)>&=& -0.7 |c\bar{c}(2S)>+\sqrt{1-0.5}|c\bar{c}g(2S)>
\nonumber \\
        |\Upsilon(3S)>&=& -0.7 |b\bar{b}(3S)>+\sqrt{1-0.5}|b\bar{b}g(3S)>
 \; ,
\eeq
with a 10\% uncertainty in the QCD sum rule estimate of the  mixing 
probabiltiy, while the $J/\Psi,\Upsilon(1S),\Upsilon(2S)$ states are 
essentially standard $q \bar{q}$ states. This solves many 
puzzles\cite{lsk09,klm11}.

  The cross sections for charmonium and bottomonium production in the
color octet model are based on the cross sections obtained from the
matrix elements for quark-antiquark and gluon-gluon octet fusion to a 
hadron H, illustrated in Fig. 1
\vspace{3cm}
\begin{figure}[ht]
\begin{center}
\epsfig{file=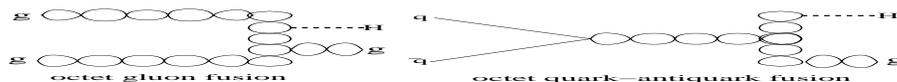,height=1.cm,width=12cm}
\caption{ Gluon and quark-antiquark color octet fusion producing hadron H}
\label{Figure 1}
\end{center}
\end{figure} 

With $q\bar{q}$ models pp cross section ratios are\cite{klm11}
$\sigma(2S)/\sigma(1S)\simeq 0.039$,$\sigma(3S)/\sigma(1S)\simeq 0.0064$. 
On the other hand, for gluonic interactions with quarks there is an
enhancement factor of $\pi^2$, for purely hybrid states, as illustrated in 
Fig. 2. For states that are approximatyely 50\% hybrid, this gives an 
enhancement factor of $\pi^2/4$, with a 10\% uncertainty, which accounts for 
the enhanced cross section ratios discussed above, in Ref\cite{klm11}, and 
below.
\newpage

\vspace{1cm}
\begin{figure}[ht]
\begin{center}
\epsfig{file=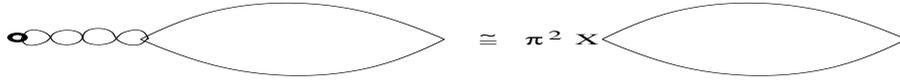,height=1.cm,width=12cm}
\caption{External field method for $\Psi'(2S)$ and $\Upsilon(3S)$ states}
\label{Figure 2}
\end{center}
\end{figure}

\section{$J/\Psi$ and $\Upsilon(1S)$ production in Cu-Cu and 
Au-Au collisions with $\sqrt{s_{pp}}$ = 200 GeV}

 The differential rapidity cross section for the production of a heavy
quark state with helicity $\lambda=0$ in the color octet model in A-A
collisions is given by

\beq
\label{1}
   \frac{d \sigma_{AA\rightarrow \Phi(\lambda=0)}}{dy} &=& 
   R^E_{AA} N^{AA}_{bin}< \frac{d \sigma_{pp\rightarrow \Phi(\lambda=0)}}{dy}>
\; ,
\eeq
where $R^E_{AA}$ is the product of the nuclear modification factor $R_{AA}$
and $S_{\Phi}$, the dissociation factor after the state $\Phi$ (a charmonium or
bottomonium state) is formed (see Ref\cite{star02}). $N^{AA}_{bin}$ is the number
of binary collisions in the AA collision, and 
$< \frac{d \sigma_{pp\rightarrow \Phi(\lambda=0)}}{dy}>$ is the
differential rapidity cross section for $\Phi$ production via nucleon-nucleon
collisions in the nuclear medium. Note that $R^E_{AA}$, which
we take as a constant, can be functions of rapidity. See 
Refs\cite{vogt08,vitov12} for a review and references to many publications.

  Experimental studies show that for $\sqrt{s_{pp}}$ = 200 GeV 
$R^E_{AA}\simeq 0.5$
both for Cu-Cu\cite{star09,phenix08} and Au-Au\cite{phenix07,star07,kks06}. The
number of binary collisions are  $N^{AA}_{bin}$=51.5 for Cu-Cu\cite{sbstar07} 
and 258 for Au-Au. The differential rapidity cross section for pp collisions 
in terms of $f_g$\cite{CTEQ6,klm11}, the gluon distribution function
($-0.8\leq y \leq 0.8$ for $\sqrt{s_{pp}}$ = 200 GeV with $f_g$ from  
Ref\cite{klm11}), is

\beq
\label{2}
     < \frac{d \sigma_{pp\rightarrow \Phi(\lambda=0)}}{dy}> &=& 
     A_\Phi \frac{1}{x(y)} f_g(\bar{x}(y),2m)f_g(a/\bar{x}(y),2m) 
\frac{dx}{dy} \; ,
\eeq  
     where $a= 4m^2/s$; with $m=1.5$  GeV for charmonium, and 5 GeV for 
bottomonium, and $A_\Phi=\frac{5 \pi^3 \alpha_s^2}{288 m^3 s}<O_8^\Phi(^1S_0)>$
\cite{klm11}. For $\sqrt{s_{pp}}$ = 200 GeV $A_\Phi=7.9 \times 10^{-4}$nb for 
$\Phi$=$J/\Psi$ and $2.13 \times 10^{-5}$nb for $\Upsilon(1S)$; $a = 
2.25 \times 10^{-4}$ for Charmonium and $2.5 \times 10^{-3}$ for Bottomium.

 The function $\bar{x}$, the effective parton x in a nucleus (A), is given in  
Refs\cite{vitov06,vitov09}:
\beq
\label{barx}
         \bar{x}(y)&=& x(y)(1+\frac{\xi_g^2(A^{1/3}-1)}{Q^2}) \nonumber \\
   x(y) &=& 0.5 \left[\frac{m}{\sqrt{s_{pp}}}(\exp{y}-\exp{(-y)})+
\sqrt{(\frac{m}{\sqrt{s_{pp}}}(\exp{y}-\exp{(-y)}))^2 +4a}\right] \;,
\eeq
with\cite{qiu04} $\xi_g^2=.12 GeV^2$. For $J/\Psi$  $Q^2=10 GeV^2$, so
$\bar{x}=1.058 x$ for Au and $\bar{x}=1.036 x$ for Cu, while for $\Upsilon(1S)$
$Q^2=100 GeV^2$, so $\bar{x}=1.006 x$ for Au and $\bar{x}=1.004 x$ for Cu.
\clearpage

From this we find the differential rapidity cross sections as shown in the 
following figures for $J/\Psi, \Psi(2S)$ and $\Upsilon(1S),\Upsilon(2S),
\Upsilon(3S)$ production via Cu-Cu and Au-Au collisions at RHIC (E=200 GeV), 
with $\Psi(2S),\Upsilon(3S)$ enhanced by $\pi^2/4$ as discussed above. The
absolute magnitudes are uncertain, and the shapes and relative magnitudes are 
our main prediction.
\vspace{2.5cm}

\begin{figure}[ht]
\begin{center}
\epsfig{file=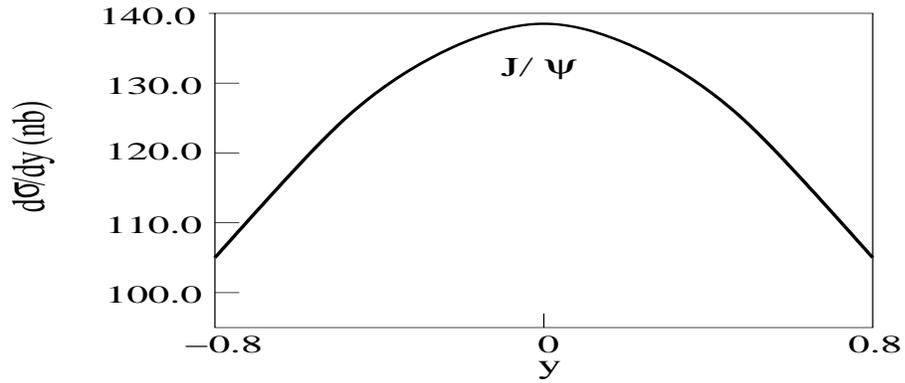,height=5 cm,width=12cm}
\caption{d$\sigma$/dy for 2m=3 GeV, E=200 GeV Cu-Cu collisions 
producing $J/\Psi$ with $\lambda=0$}
\label{Figure 3}
\end{center}
\end{figure}
\vspace{2.5cm}

\begin{figure}[ht]
\begin{center}
\epsfig{file=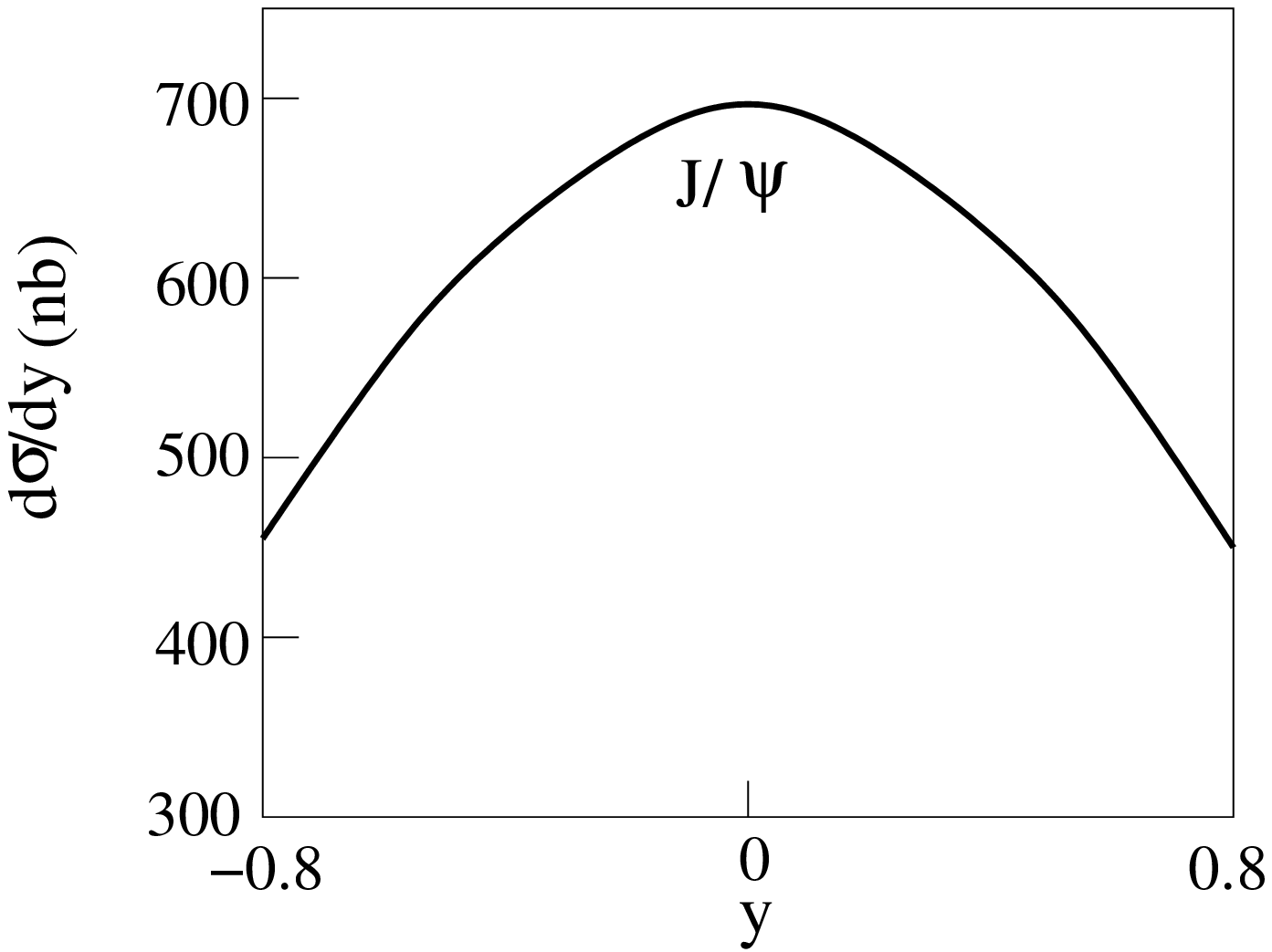,height=5 cm,width=12cm}
\caption{d$\sigma$/dy for 2m=3 GeV, E=200 GeV Au-Au collisions 
producing $J/\Psi$ with $\lambda=0$}
\label{Figure 4}
\end{center}
\end{figure}
\clearpage

\begin{figure}[ht]
\begin{center}
\epsfig{file=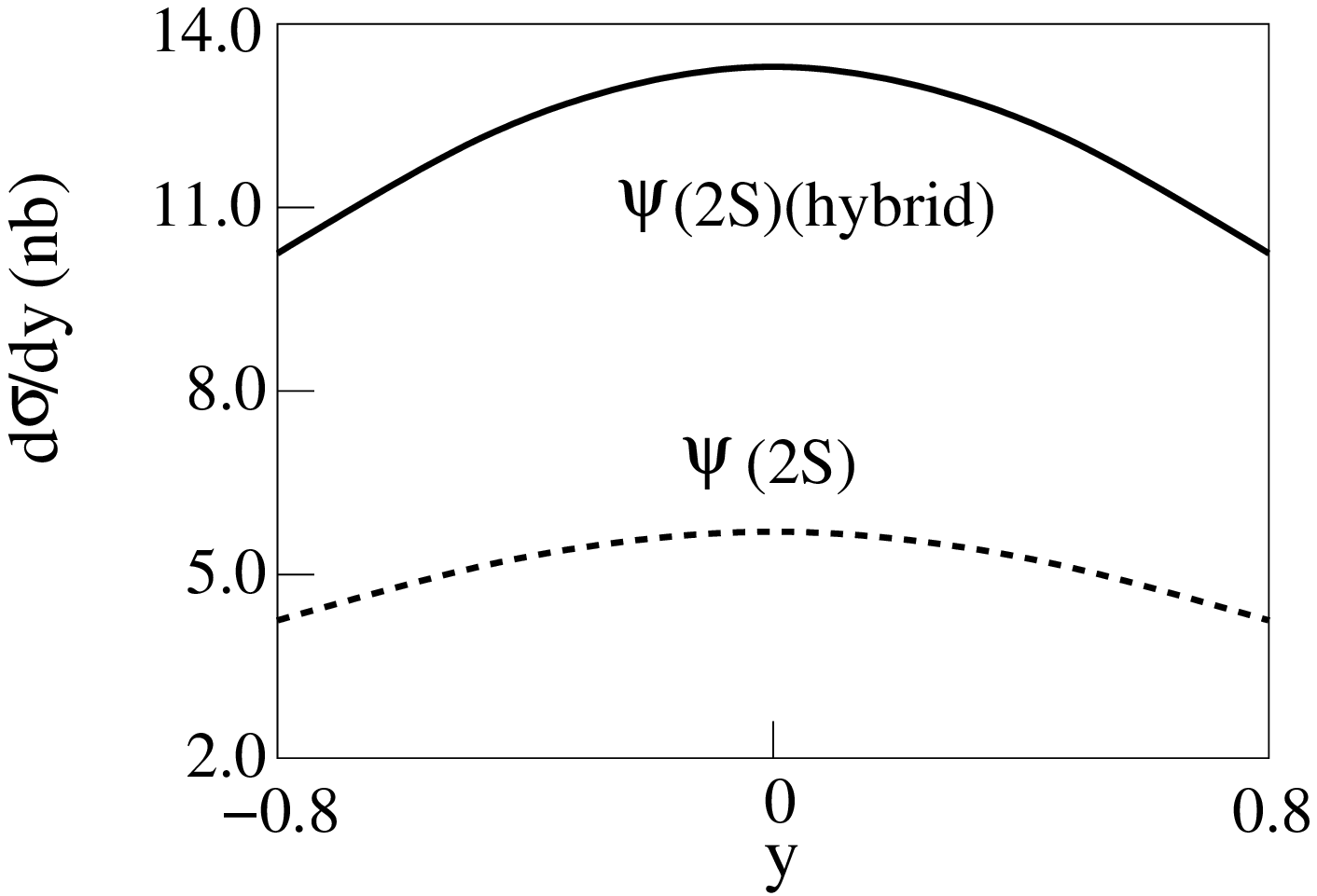,height=6 cm,width=12cm}
\caption{d$\sigma$/dy for 2m=3 GeV, E=200 GeV Cu-Cu collisions 
producing $\Psi(2S)$ with $\lambda=0$. The dashed curve is for the standard
$c\bar{c}$ model.}
\label{Figure 5}
\end{center}
\end{figure}

\vspace{4cm}

\begin{figure}[ht]
\begin{center}
\epsfig{file=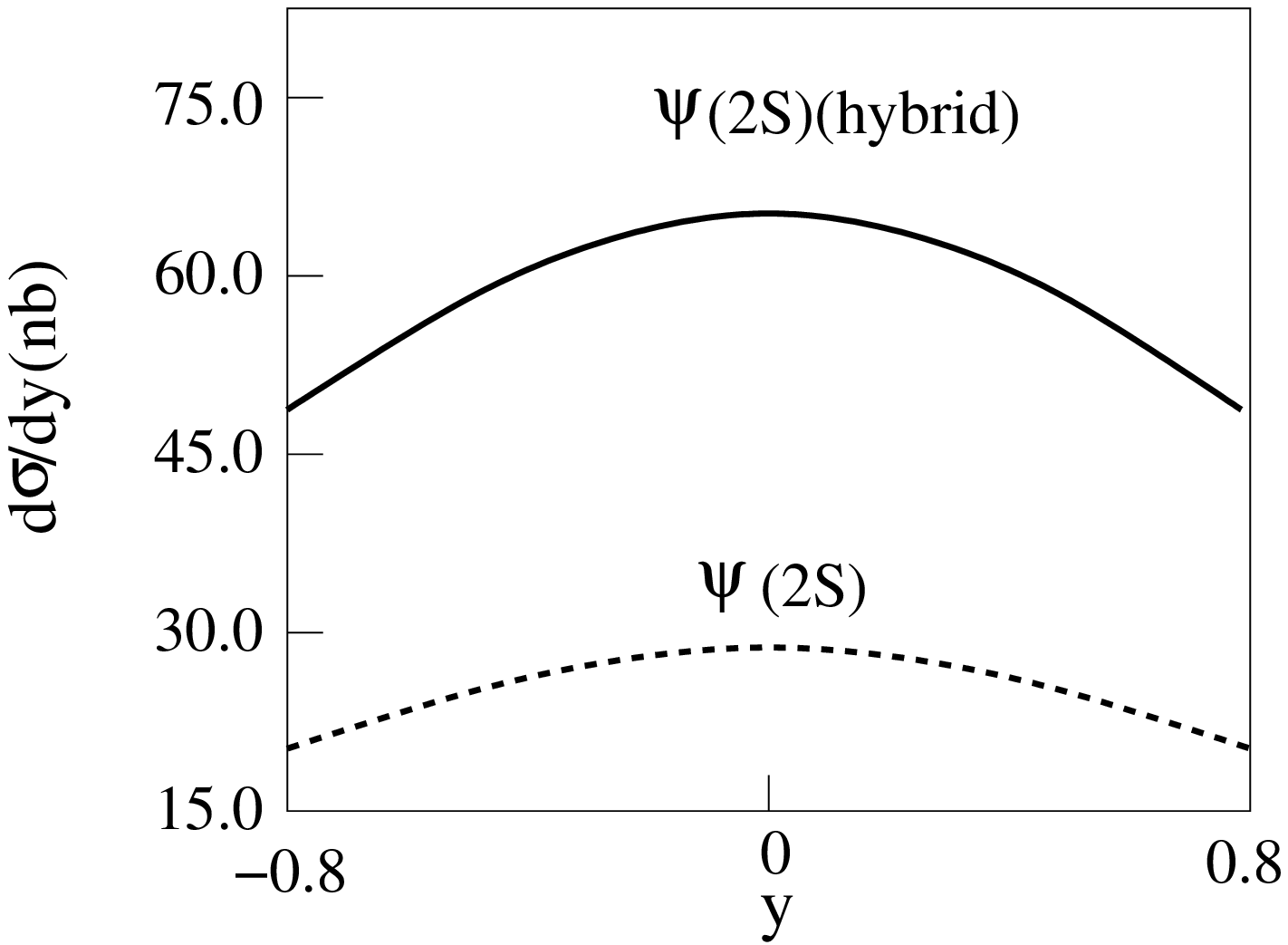,height=6 cm,width=12cm}
\caption{d$\sigma$/dy for 2m=3 GeV, E=200 GeV Au-Au collisions 
producing $\Psi(2S)$ with $\lambda=0$. The dashed curve is for the standard
$c\bar{c}$ model.}
\label{Figure 6}
\end{center}
\end{figure}
\clearpage

\begin{figure}[ht]
\begin{center}
\epsfig{file=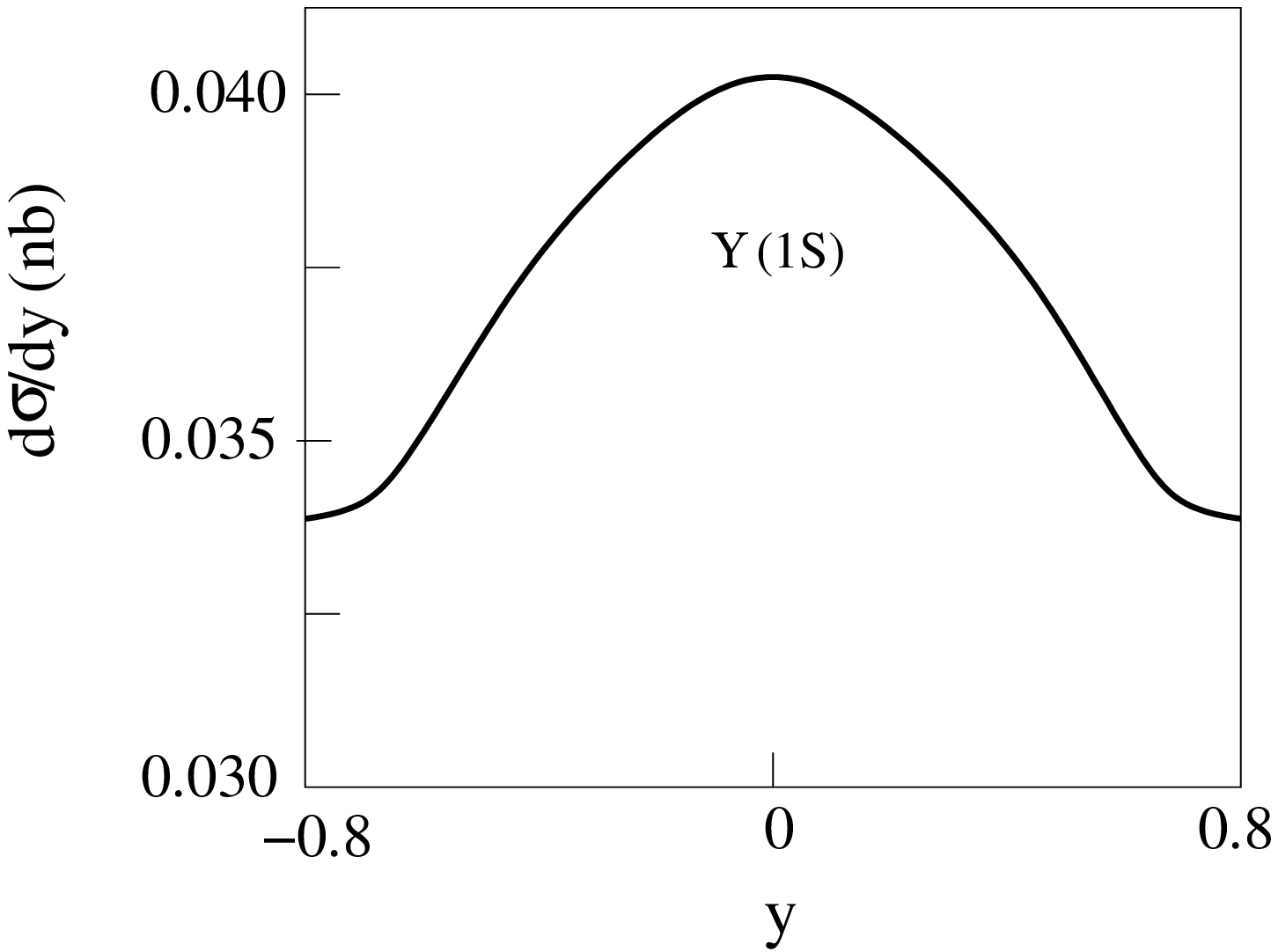,height=7 cm,width=12cm}
\caption{d$\sigma$/dy for 2m=10 GeV, E=200 GeV Cu-Cu collisions 
producing $\Upsilon(1S)$ with $\lambda=0$}
\label{Figure 7}
\end{center}
\end{figure}
\vspace{3 cm}

\begin{figure}[ht]
\begin{center}
\epsfig{file=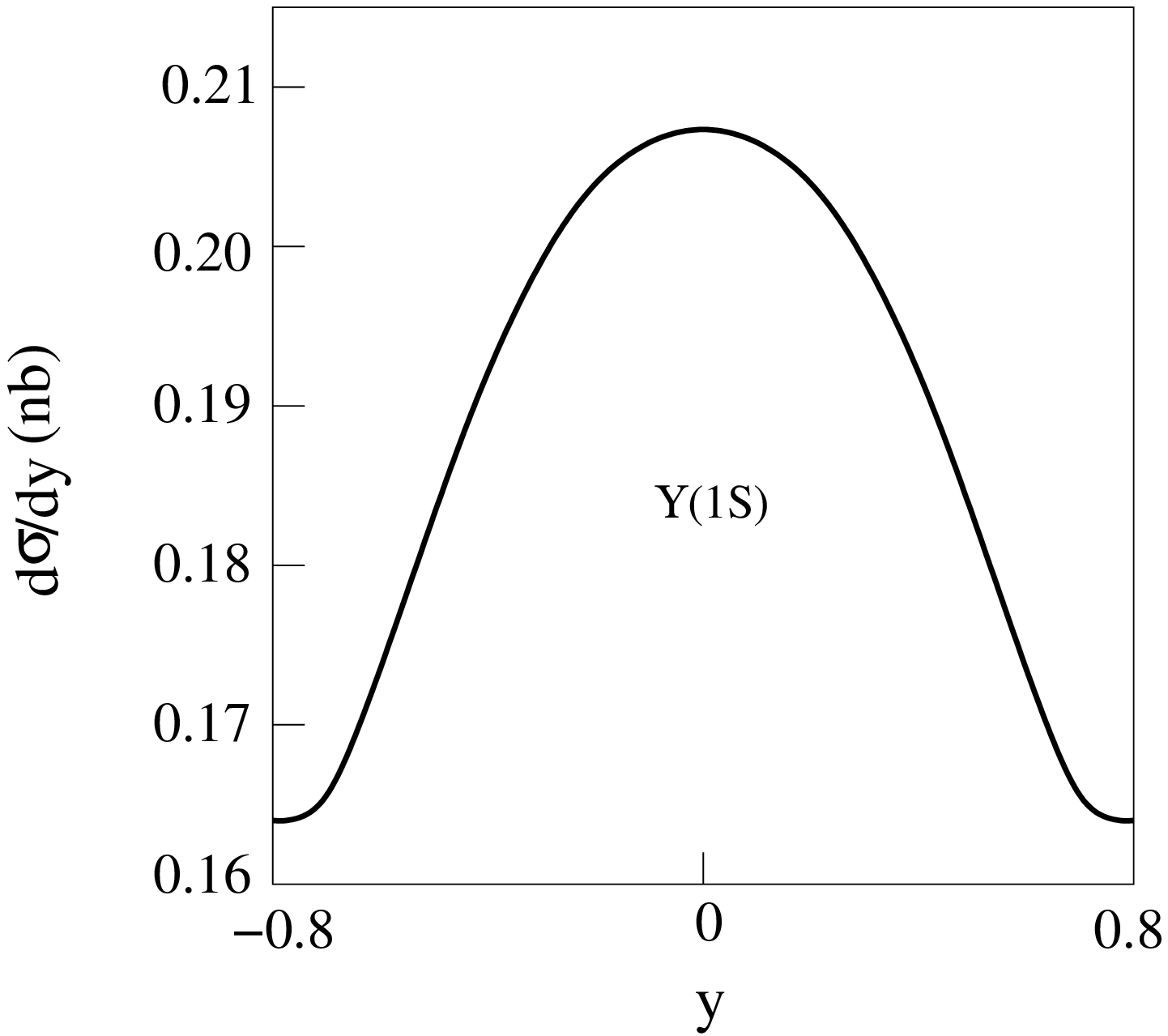,height=6 cm,width=12cm}
\caption{d$\sigma$/dy for 2m=10 GeV, E=200 GeV Au-Au collisions 
producing $\Upsilon(1S)$ with $\lambda=0$}
\label{Figure 8}
\end{center}
\end{figure}
\clearpage

\begin{figure}[ht]
\begin{center}
\epsfig{file=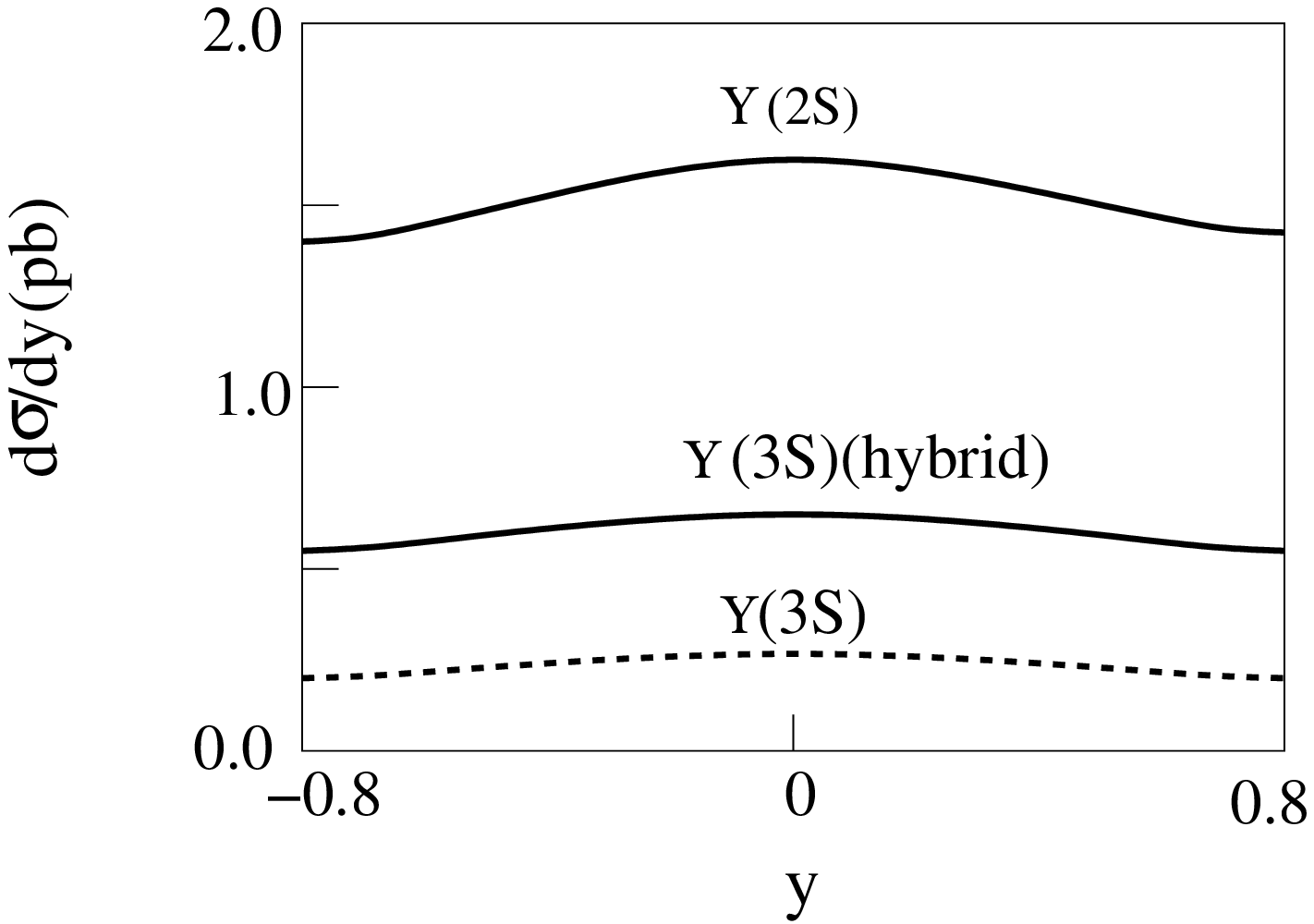,height=7 cm,width=12cm}
\caption{d$\sigma$/dy for 2m=10 GeV, E=200 GeV Cu-Cu collisions 
producing $\Upsilon(2S),\Upsilon(3S)$ with $\lambda=0$. For $\Upsilon(3S)$
the dashed curve is for the standard $b\bar{b}$ model.}
\label{Figure 9}
\end{center}
\end{figure}
\vspace{5 cm}

\begin{figure}[ht]
\begin{center}
\epsfig{file=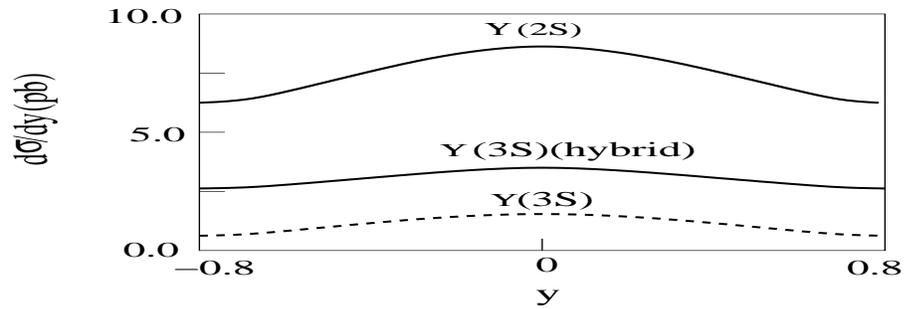,height=4 cm,width=12cm}
\caption{d$\sigma$/dy for 2m=10 GeV, E=200 GeV Au-Au collisions 
producing $\Upsilon(2S),\Upsilon(3S)$ with $\lambda=0$. For $\Upsilon(3S)$
the dashed curve is for the standard $b\bar{b}$ model.}
\label{Figure 10}
\end{center}
\end{figure}
\newpage

\section{Ratio of $\Psi'(2S)$ to $J/\Psi$ cross sections}

In this section we discuss the ratios of the charmonium cross sections
for p-p and A-A collisions at RHIC. In order to estimate the $\Psi'(2S)$
to $J/\Psi$ ratios in A-A collisions we make use of recent experimental
results on $\Upsilon(mS)$ state production at the LHC.

\subsection{Ratios for p-p collisions} 

  In Ref\cite{klm11} we discussed the $\Upsilon(mS)$ cross section ratios,
showing that the error in the ratios is small as it is given by the wave 
functions for the standard model and the enhancement factor of 
$(1 \pm .1)\times \pi^2/4$ for the mixed hybrid, as discussed in Section 2. 
Now there are accurate 
measurements of the $\Psi'(2S)$ to $J/\Psi$ ratio for p-p production at RHIC. 
From the standard (st), hybrid model(hy) one finds for p-p production of 
$\Psi'(2S)$ and $J/\Psi$
\beq
\label{ppratio}
    \sigma(\Psi'(2S))/\sigma(J/\Psi(1S))|_{st} &\simeq& 0.27 \nonumber \\
    \sigma(\Psi'(2S))/\sigma(J/\Psi(1S))|_{hy} &\simeq& 0.67\pm 0.07 \; ,
\eeq
while the PHENIX experimental result for the ratio\cite{phoenix} $\simeq$
0.59. Therefore, as in our earlier work the hybrid model is consistent with
experiment, while the standard model ratio is too small.
 
\subsection{Ratios for Pb-Pb collisions}

  The recent CMS/LHC result comparing Pb-Pb to p-p Upsilon
production\cite{cms2-11} found
\beq
\label{CMS2}
     [\frac{\Upsilon(2S) +\Upsilon(3S)}{\Upsilon(1S)}]_{Pb-Pb}/
    [\frac{\Upsilon(2S) +\Upsilon(3S)}{\Upsilon(1S)}]_{p-p} &\simeq& 
0.31^{+.19}_{-.15} \pm .013(syst) \; ,
\eeq
while in our previous work on $p-p$ collisions we found the ratio 
$\sigma(\Upsilon(3S))/\sigma(\Upsilon(1S))|_{p-p}$
of the standard $|b\bar{b}>$ model was $4/\pi^2 \simeq 0.4$ of the hybrid
model. This suggests a suppression factor for $\sigma(b\bar{b}(3S))/
\sigma(b\bar{b}(1S))$, or $\sigma(c\bar{c}(2S)/\sigma(c\bar{c}(1S))$ of 
0.31/.4 as these components travel 
through the QGP; or an additional factor of 0.78 for $\Psi'(2S)$ to $J/\Psi$
production for $A-A$ vs $p-p$ collisions. Therefore from Eq(\ref{ppratio}) 
one obtains our estimate using our mixed hybrid theory for this ratio
\beq
  \sigma(\Psi'(2S))/\sigma(J/\Psi(1S))|_{A-A{\rm \; collisions}} &\simeq& 
0.52 \pm 0.05 
\eeq

\section{Ratio of $\Upsilon(2S)$ and $\Upsilon(3S)$
 to $\Upsilon(1S)$ cross sections}

  In our previous work\cite{klm11} we estimated the ratios of $\Upsilon(2S)$ 
and $\Upsilon(3S)$ to $\Upsilon(1S)$ cross sections in comparison with an
experiment published in 1991\cite{fermi91}. Our result for p-p collisions, with
uncertainty due to separating $\Upsilon(2S)$ from $\Upsilon(3S)$, was
\beq
\label{lsk11}
     \Upsilon(3S)/\Upsilon(1S)|_{p-p} &\simeq& 0.14-0.22 \; ,
\eeq
for our mixed hybrid theory, while the standard model would give
$\frac{\Upsilon(3S)}{\Upsilon(1S)} \simeq 0.06$. A recent CMS 
result\cite{cms12}, with a correction factor for acceptance and 
efficiency of the $\Upsilon(3S)$ to the  $\Upsilon(1S)$ state, which was 
estimated to be approximately 0.29, was found to be
\beq
\label{cms12}
     \Upsilon(3S)/\Upsilon(1S)|_{p-p} &\simeq& 0.12 \; ,
\eeq
with the mixed hybrid theory in agreement within errors, while the standard 
model differs by a factor of two.

  The new CMS experiment's main objective\cite{cms12} is to test for 
$\Upsilon$ suppression in PbPb collisions, with estimates of the following
quantities:
\beq
\label{cmsz12}
  && \frac{[\Upsilon(2S)/\Upsilon(1S)]_{PbPb}}
 {[\Upsilon(2S)/\Upsilon(1S)]_{pp}} \nonumber \\
  &&\frac{[\Upsilon(3S)/\Upsilon(1S)]_{PbPb}}
 {[\Upsilon(3S)/\Upsilon(1S)]_{pp}} \; .
\eeq
  The studies of AA collisions for Bottomonium states, which cannot be
carried out at RHIC but are an important part of the LHC CMS program,
will be carried out in our future research.

\section{Conclusions}

  We have studied the differential rapidity cross sections for
$J/\Psi, \Psi(2S)$ and $\Upsilon(nS)(n=1,2,3)$ production via Cu-Cu and
Au-Au collisions at RHIC (E=200 GeV) using $R^E_{AA}$, the product of the 
nuclear modification factor $R_{AA}$ and the dissociation factor $S_{\Phi}$,
$N^{AA}_{bin}$ the binary collision number, and the gluon distribution
functions from previous publications. This should give some guidance for 
future RHIC experiments, although at the present time the $\Upsilon(nS)$ 
states cannot be resolved.

  The ratio of the production of $\sigma(\Psi'(2S))$, which in our mixed
hybrid theory is 50\% $c\bar{c}(2S)$ and 50\% $c\bar{c}g(2S)$ with a 
$10\%$ uncertainty, to
$J/\Psi(1S)$, which is the standard $c\bar{c}(1S)$, will be an important
test of the production of the quark-gluon plasma. Using the hybrid model
and suppression factors from previous theoretical estimates and experiments
on $\Upsilon(mS)$ state production at the LHC, we estimate that the ratio 
of $\Psi'(2S)$ to $J/\Psi(1S)$ production at RHIC via A-A collisions will 
be about $0.52\pm 0.05$.

\vspace{1cm}
\Large{{\bf Acknowledgements}}\\
\normalsize
This work was supported in part by a grant from the Pittsburgh Foundation,
and in part by the DOE contracts W-7405-ENG-36 and DE-FG02-97ER41014.
We thank Drs. Ramona Vogt and Ivan Vitev for helpful discussions and
suggestions.

\end{document}